\documentclass[conference]{IEEEtran}
\IEEEoverridecommandlockouts
\usepackage{cite}
\usepackage{amsmath,amssymb,amsfonts}
\usepackage{textcomp}
\usepackage{xcolor}
\def\BibTeX{{\rm B\kern-.05em{\sc i\kern-.025em b}\kern-.08em
    T\kern-.1667em\lower.7ex\hbox{E}\kern-.125emX}}
    
\usepackage[utf8]{inputenc} 
\usepackage[T1]{fontenc}    
\usepackage{hyperref}       
\usepackage{url}            
\usepackage{booktabs}       
\usepackage{nicefrac}       
\usepackage{microtype}      
\usepackage{graphicx}
\usepackage{wrapfig}
\usepackage[ruled,vlined]{algorithm2e}
\usepackage{newfloat}

\usepackage{algpseudocode}
\begin{document}

\title{A New Parallel Algorithm for Sinkhorn Word-Movers Distance Using Fused SDDMM-SpMM on PIUMA and Xeon CPU}

\author{\IEEEauthorblockN{Jesmin Jahan Tithi}
\IEEEauthorblockA{\textit{Parallel Computing Labs} \\
\textit{Intel}\\
Santa Clara, CA, USA \\
jesmin.jahan.tithi@intel.com}
\and
\IEEEauthorblockN{Fabrizio Petrini}
\IEEEauthorblockA{\textit{Parallel Computing Labs} \\
\textit{Intel}\\
Santa Clara, CA, USA \\
fabrizio.petrini@intel.com}
}
\newcommand{\PIUMA}[1]{\textcolor{black}{#1}}
\newcommand{\IMP}[1]{\textcolor{black}{#1}}
\newcommand{\new}[1]{\textcolor{black}{#1}}

\maketitle
\begin{abstract}
The Word Movers Distance (WMD) measures the semantic dissimilarity between two text documents by computing the cost of optimally moving all words of a source/query document to the most similar words of a target document. Computing WMD between two documents is costly because it requires solving an \(O(V^3log(V))\) optimization problem where \(V\) is the number of unique words in the document. Fortunately, WMD can be framed as an Earth Mover's Distance (EMD) for which the algorithmic complexity can be reduced to \(O(V^2)\) by adding an entropy penalty to the optimization problem and solving it using the Sinkhorn-Knopp algorithm. Additionally, the computation can be made highly parallel by computing the WMD of a single query document against multiple target documents at once, for example by finding whether a given tweet is similar to any other tweets of a given day. 

Sinkhorn WMD is a key kernel used in many ML/NLP applications and usually gets implemented in Python. However, a naive Python implementation may leave 1000x performance on the table for Sinkhorn WMD even though it may internally call optimized C++ BLAS routines. We present a new shared-memory parallel Sinkhorn-Knopp algorithm to compute the WMD of one document against many other documents by adopting the \(O(V^2)\) EMD algorithm. We algorithmically transform \(O(V^2)\) dense compute-heavy EMD version into an equivalent sparse one using new fused SDDMM-SpMM (sparse selection of dense-dense matrix-, sparse-dense matrix- multiplication) kernels. We implemented and optimized this algorithm for two very different architectures --- the new Intel Programmable Integrated Unified Memory Architecture (PIUMA) and Intel Xeon CPUs. We show that we were able to reach close to peak performance on both platforms.
\end{abstract}




\section{Introduction}
The Word Movers Distance (WMD)~\cite{WMD} measures dissimilarity between two text documents as the minimum distance that the words of one document need to ``move/travel" to reach the words of another document. To represent a word, WMD uses the word-embedding vector from word2Vec~\cite{word2vec}, encoding semantic meanings of the words from their local co-occurrences in sentences learned via neural network. 
The WMD is used in many text analysis applications such as: agglomerative short text clustering \cite{10.1007/978-3-030-14799-0_11}, evaluations of generated texts, for document retrieval\cite{brokos2016using}, for machine translation \cite{zhang2016building}, to create document embedding \cite{wu2018word}, and automatic essay evaluations \cite{tashu2018pair}. 
Although the concept of WMD was introduced in 2015, a search in Google Scholar with the keyword "Word Mover's Distance" brings up 1,460 results suggesting its growing popularity and usage. \new{WMD computation is a performance-critical component for many large-scale ML/NLP applications that often process terabytes of data and would require HPC resources for processing. Therefore, accelerating computation of WMD is important.} 

Computing WMD between two documents requires solving an optimization problem that costs \(O(V^3log(V))\)\footnote{The cubic cost can be linked to the cost of the network flow problem.} where \(V\) is the number of unique words in the documents\cite{pele2009fast}. This cost can be reduced by following an idea presented in \cite{Cuturi:2013:SDL:2999792.2999868} that approximates the Earth Mover's Distance (EMD) and that can be adopted to WMD. This algorithm uses an entropy penalty to the optimization problem that encourages the solution to lie `close' to a transportation plan that sends equal mass from each point in the source document to each point in the target document. It then uses the Sinkhorn-Knopp matrix scaling algorithm \cite{Knight:2008:SAC:1404637.1404647} to solve the optimization problem which takes \(O(V^2)\) time. Additionally, this computation can be made highly parallel by computing the WMD of a single query/source document against multiple target documents (batch) at once. A practical use case of this is to find whether a tweet is similar to any other tweets in a given day. We use the above optimization ideas to develop an efficient parallel WMD algorithm that requires \(O(V^2/p)\) time per query where \(V\) is the number of words in the vocabulary and \(p\) is the number of threads used.

A naive implementation of the WMD algorithm may make it a dense kernel-heavy (i.e., dense portion of the kernel takes major fraction of the runtime) workload. However, we show that by using an improved algorithmic design, it can be transformed to a sparse kernel-heavy workload which makes it faster by pruning unnecessary work, storage and bandwidth cost. 

We also show an early preview of this algorithm's performance on the new Intel Programmable Integrated Unified Memory Architecture (PIUMA). PIUMA is optimized for irregular and sparse workloads and developed under the DARPA HIVE \cite{Hive2020piuma} program. PIUMA natively supports a Distributed Global Address Space (DGAS) where any memory location can be read/written directly using normal load/store operation significantly easing programming. 
\new{PIUMA implements native scale-out in DGAS with many cores and memory pools}.\footnote{The PIUMA hardware is going through the power-on phase. The results presented in this paper are based on Intel's Sniper simulator with PIUMA's RTL validated on a large-scale FPGA cluster. We expect to validate the experimental results on real HW within the next few months.}  Hence we simulated WMD's performance on PIUMA and expect that it would strongly scale up to \(4096\) cores (\(262,144\) threads) on PIUMA whereas on Xeon CPU the scalability starts flattening much earlier.

{\bf{Contributions:}}
Our contributions are as follows:
\begin{itemize}

\item {\bf Algorithmic Innovation:} We present a sparse algorithm with better asymptotic bounds to compute the Word Movers Distances of one source document to many target documents. We propose a new kernel called SDDMM\_SpMM fusing the Sparse x Dense x Dense matrix multiplication (SDDMM) and the Sparse x Dense matrix multiplication (SpMM) kernels and used it in the Sinkhorn-Knopp WMD algorithm. To our knowledge, this is the first paper showing SDDMM\_SpMM and its application to WMD.\footnote{An archive version is available whose reference has been omitted for double-blind purposes. A latter research also proposes the fusion of SDDMM\_SpMM in a generic sense but does not appear to cover the case with three dense and two sparse inputs that we encounter for Sinkhorn.}

\item {\bf Parallelization:} We present a shared-memory parallel algorithm to compute the WMD of one document against many documents at once by combining a fast EMD  \cite{Cuturi:2013:SDL:2999792.2999868} and the Sinkhorn-Knopp matrix-scaling algorithm~\cite{Knight:2008:SAC:1404637.1404647}. We show implementations of this algorithm on the PIUMA and Xeon CPU architectures.

\item {\bf Theoretical Analysis:} We show the theoretical runtime analysis of the presented algorithm.

\item {\bf Porting to PIUMA:} The presented WMD algorithm is a mix of sparse and dense compute. We show how to port this algorithm on the new PIUMA Architecture and discuss the performance implications of PIUMA features.PIUMA is optimized for sparse and irregular workloads and the first version of PIUMA is not equipped with a dense accelerator. PIUMA cores are scalar and have low (and balanced) `flops per byte' capacity unlike Xeon CPUs (or GPUs). 

\item {\bf Experimental Analysis:} We show that on state-of-the-art Xeon CPUs, the parallel implementation of this new algorithm is \(\approx 700\times\) faster than the parallel Python code that internally uses parallel math kernel libraries and \(42\times\) faster than a baseline C++ implementation. The parallel implementation achieves \(27\times\) speedup on \(56\) cores sharing resources across \(2\) NUMA sockets of an Intel Cascade Lake machine w.r.t. its sequential run which is close to ideal for a bandwidth bound kernel.  
\end{itemize}

\section{Background}
In this section, we discuss the necessary background and some prior research that make the foundation for our work.

\subsection{Prior Work}
The WMD attempts to capture semantic similarity between documents. With words represented as vectors, each text document can be considered as a weighted point cloud of embedded words. Then, the distance between two text documents, say, \(A\) and \(B\) would be the minimum cumulative distance that words from document \(A\) need to travel to match exactly the point cloud of document \(B\). Distance between two words \(i\), \(j\) can be measured by the euclidean distance, \(m(i, j)\) = \({\parallel x_i - x_j \parallel}^2\) = \(\sqrt{{\sum{{\left(embedding[i] - embedding[j]\right)}^2}}}\) between the high-dimensional vectors \(x_i\) and \(x_j\) associated with each word's word2vec representation. Essentially, \(m(i, j)\) gives the moving cost of word \(i\) to word \(j\) and a small distance indicates that words are closely semantically related.   
 
 \begin{figure}[t]
  \centering
  \includegraphics[width=0.4\textwidth]{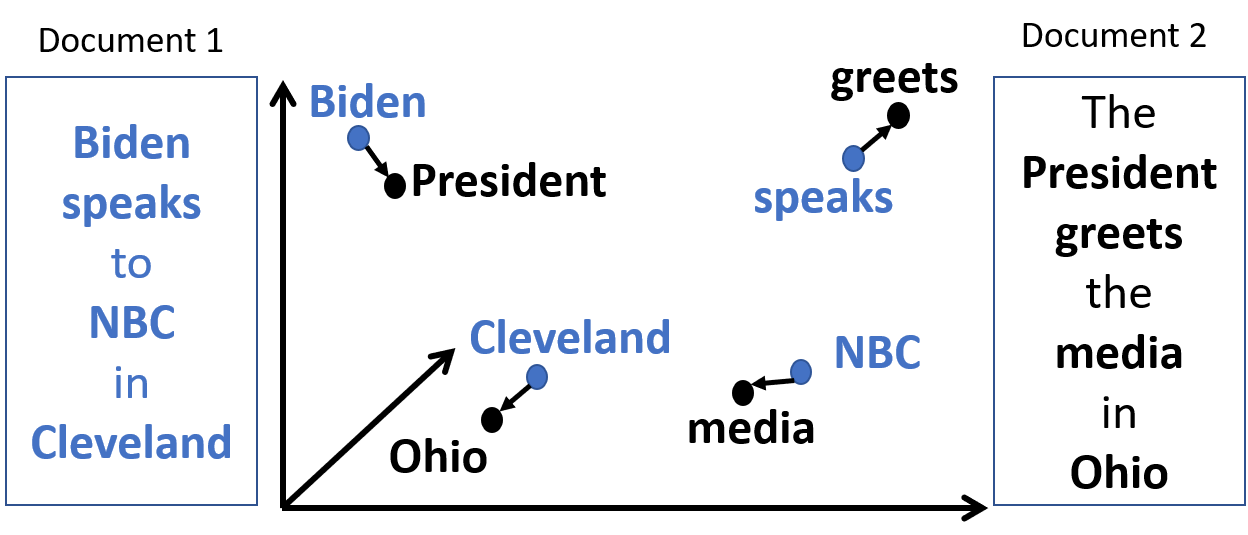}
  \caption{Example of a WMD calculation between two single-sentence documents.}
  \label{fig:WMD-principle}
\end{figure}
Suppose, we have two documents \(A\) and \(B\) where A = ``Biden speaks to the NBC in Cleveland'', and B = ``The President greets the media in Ohio''. These two sentences have intuitively the same meaning and therefore, these sentences should be ``more similar" to each other than, for example, to the sentence ``Biden praised Bangladesh at it's 50th Independence Day''. After throwing away the information about word order, capitalization and removing the frequent and uninformative stop-words (e.g., in, to, the), we get the following ``bag-of-words'' representation of \(A\) and \(B\): A = [`cleveland', `nbc', `speaks', `biden'] and B = [`ohio', `greets', `president', `media']. Since \(A\) and \(B\) don't contain any of the same words, one can not look at the set intersection to measure sentence similarity. However, if the words are represented as vectors in the word-embedding space, it is expected that the word `biden' would be close to `president' and `Cleveland' will be close to `Ohio' and \(m(media, nbc) < m(media, biden)\), where,
m(media, nbc) = \(\sqrt{{\sum{{\left(embedding[media] - embedding[nbc]\right)}^2}}}\). 
Even though there is no overlap in words between the two documents, the word embedding provides the semantic association between disjoint words. 

Figure \ref{fig:WMD-principle} shows the WMD between documents \(A\) and \(B\) as the minimum cumulative distance that all non-filler words in the document \(1\) (blue) need to travel to exactly match the words in document \(2\) (black). Here, the WMD is the sum of the lengths of all black arrows.

The WMD is similar to Earth Movers Distance (EMD) which measures the cost of the optimal way to transport dirt from a set of source piles to a set of destination piles. EMD represents a family of well-studied problems in operations research for which several specialized solvers~\cite{pele2009fast} have been developed. If the word embedding vectors of size \(w\) are considered as points in a \(w\)-dimensional space, then the distance between words \(i\) and \(j\) can be interpreted as the `cost' of transporting a unit of `mass' from point \(embedding[i]\) to point \(embedding[j]\). Then, sentences \(A\) and \(B\) can be considered assets of points in that space, with a unit of mass piled on each point in \(A\) and \(B\). With this setting, the Word Mover's Distance (WMD) would be the minimum cumulative cost of transporting all of the mass from points [\(embedding[i]\) for \(i\) in \(A\)] to points [\(embeddings[j]\) for \(j\) in \(B\)]. Note that mass can and might flow from a single point in \(A\) to multiple points in \(B\) and vice-versa, because, words can, of course, occur multiple times and can be mapped to different related words. The \(num\_words(A)\) \(\times\) \(num\_words(B)\) matrix that specifies the mass flow cost from \(A\) to \(B\) is called the `transportation cost' matrix. 
 
\begin{figure*}[t]
	\centering
	\includegraphics[width=0.8\textwidth]{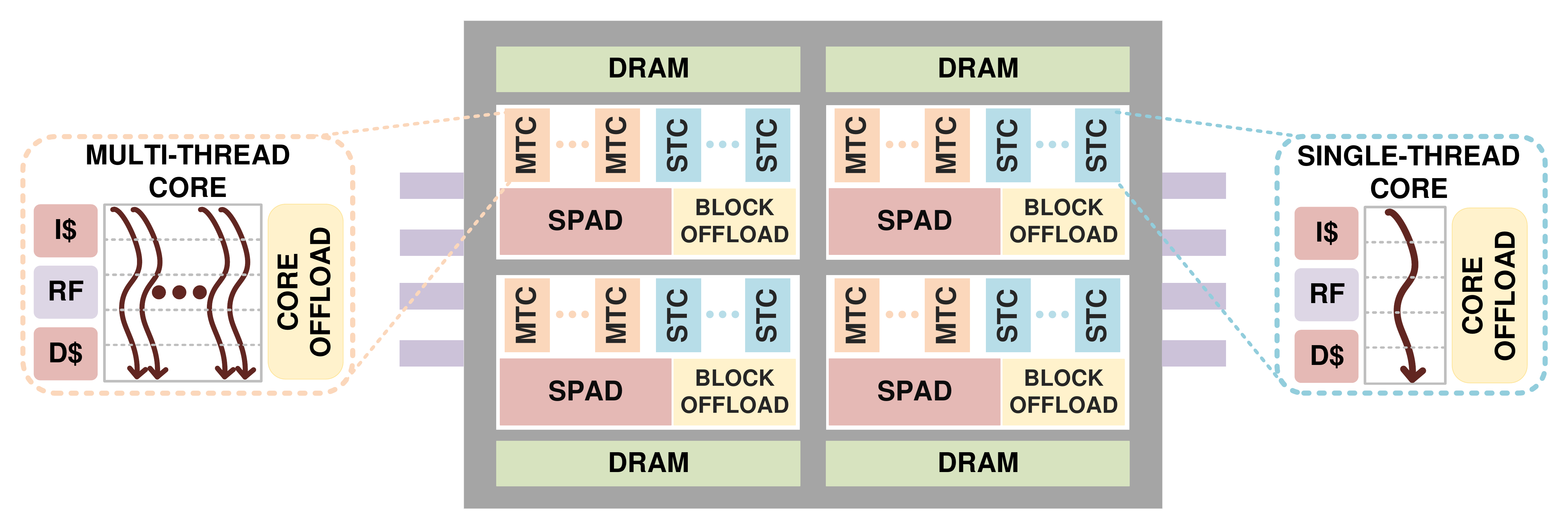}
	\caption{High-level diagram of PIUMA architecture (collected from \cite{Hive2020piuma}).}
	\label{fig:PIUMA_arch_overview}
\end{figure*}

If we have a database of \(5\)\,M documents, running a single query to compute the WMD of document \(A\) to all other \(5\)\,M documents would involve solving \(5\)\,M optimization problems of size \(O(V^3log(V))\) where \(V\) is the number of unique words (e.g., 100\,K) in the documents, then it would be quite costly. The original paper by Kushner et. al. ~\cite{WMD} used a flow-based approach to balance the total `in' and `out' flows of masses to compute the WMD. Many recent works follow a flow-based technique to compute WMD ~\cite{10.1007/978-3-030-14799-0_11}, ~\cite{brokos2016using}, ~\cite{zhang2016building}, ~\cite{wu2018word}, ~\cite{tashu2018pair} which may take \(O(V^3log(V))\) time in the worst case following the cost of max flow min cost problem \cite{orlin1993faster}. Several pruning ideas have been proposed in \cite{WMD} to speed up the document retrieval process that reduces the number of expensive WMD evaluations per query document. However, the individual WMD computation remains prohibitively expensive in practice. 

Cuturi et. al. ~\cite{Cuturi:2013:SDL:2999792.2999868} proposed an approximation to the Optimal Transportation Distance that reduced the cost per query for the earth movers distance (EMD) problem to \(O(V^2)\) by adding an entropy regularization term: basically, the optimization problem is regularized with this entropy term following the maximum-entropy principle. This encourages the solution to lie `close' to the (trivial) transportation plan that sends equal mass from each point in \(A\) to each point in \(B\). `Closeness' is measured by Kullback-Leibler (KL)-divergence - a measure of how one probability distribution (e.g., word frequency of one document) is different from a second reference probability distribution. Using this entropy term, the optimization problem is converted from a general linear programming (LP) problem into a convex problem, which can be solved using the Sinkhorn-Knopp matrix scaling algorithm ~\cite{Knight:2008:SAC:1404637.1404647}. The authors in ~\cite{Cuturi:2013:SDL:2999792.2999868} called the new approximated distance as the Sinkhorn distance and proved that for large enough entropy, the Sinkhorn distance is equivalent to the optimal transportation distance (proof is shown in  ~\cite{Cuturi:2013:SDL:2999792.2999868}). Furthermore, the Sinkhorn distance is symmetric and satisfies triangle inequality and is a metric. The added penalty/entropy makes the problem easier to solve, provides asymptotic cost reduction and one can use the fast Sinkhorn-Knopp matrix scaling algorithm which is also easily parallelizable. This formulation makes the problem a non-flow like problem. In this paper, we adopt the above approximation ideas to optimize computation of WMD \cite{WMD}. A very recent paper \cite{Li:2019:CES:3308558.3313397} used this approach as part of their work on short-text clustering. \new{However, they used a sequential and the naive dense approach} to compute WMD. We also found their algorithmic description complicated to apprehend. 

\IMP{
In our optimized WMD algorithm, we used a new SDDMM\_SpMM Kernel to make the algorithm a sparse-\new{compute} heavy one. Although SDDMM and SPMM \cite{hong2019adaptive,gale2020sparse} kernels are very popular and their fusion might be intuitive, to the best of our knowledge SDDMM\_SpMM\footnote{proposed in an archive article published in May 2020 and reference omitted for blind review purpose} could be considered a contribution of this paper. A latter research \cite{rahman2020fusedmm} called it FusedMM and created a general purpose library targeting Graph Neural Network and similar class of applications. However, this generic class fails to capture all SDDMM\_SpMM input patterns required in Sinkhorn-WMD as discussed later.
}

\subsection{PIUMA}
\label{PIUMAArch}

In this section, we give a high-level description of the PIUMA Architecture. Details can be found in the original article \cite{Hive2020piuma}. \PIUMA{Here, we will be highlighting the features of PIUMA that are leveraged by the WMD implementation and contrast them to a Xeon CPU.}

\PIUMA{
Unlike traditional homogeneous architectures, the PIUMA architecture \cite{PIUMABlog,HIVEWiki,Hive2020piuma} consists of two different types of cores, Multi-threaded core (MTC) and Single-threaded core (STC). The MTCs are round-robin-multithreaded scalar cores with 64 threads and each thread can have one instruction on the flight. MTCs are usually used for the data-parallel portion of a code. STCs are single-threaded in-order stall-on-use cores that can exploit some instruction and memory-level parallelism and are usually used for single-thread performance-sensitive tasks, such as memory allocations and thread management \cite{Hive2020piuma}.} We use the STCs for the data reading and memory allocations and MTCs to work on the core kernel (a.k.a actual computations).

\PIUMA{As Figure \ref{fig:PIUMA_arch_overview} shows, each MTC and STC have a local instruction cache (I\$), data cache (D\$), and register file (RF) \cite{Hive2020piuma}. PIUMA allows selectively caching any data through the use of a unique bit in the address space. To facilitate scalability, coherency in caches is not maintained across the whole system.} This is different than standard Xeon CPUs where all data is cached by default and there is usually a global coherency. For WMD, we selectively cached read-only data that would likely have spatial and temporal reuse.

\PIUMA{MTCs and STCs are grouped into blocks that have large low-latency scratchpad (SPAD) storage. Each block has offload engines to efficiently fetch large chunks of data to the intended storage \cite{Hive2020piuma}. Each core's offload region in Figure~\ref{fig:PIUMA_arch_overview} 
contains a direct memory access (DMA) engine that executes gather, scatter, copy, initialization, reduction, and broadcast operations \cite{Hive2020piuma}. ``Programmers select which memory accesses to cache (e.g., local stack), which to put on SPAD (e.g., frequently reused data structures or the result of a DMA gather operation), and which to store in the global address space" \cite{Hive2020piuma}.} For WMD, we used SPAD storage for arrays that require atomic accesses since \PIUMA{atomics on SPAD are faster than main memory}. Other matrices are allocated to the distributed global address space. We used DMAs for data initialization and copies.

"PIUMA implements hardware distributed global address space (DGAS), which enables each core to uniformly access memory across the full system with one address space" \cite{Hive2020piuma}. "Address translation tables contain programmable rules to translate application
memory addresses to physical locations, to arrange the address
space to the need of the application (e.g., address striped/interleaved, block
partitioned, etc.)"\cite{Hive2020piuma}. The memory controllers (one per block) can support native 8-byte accesses and standard cache line accesses as well. Again, this is different from conventional architectures which supports usually only cache line (or larger) transfer sizes. For WMD, we used the default striped allocation to distribute the data evenly across the memory controllers which evens out the access pressures. We benefit from the 8-byte accesses while accessing sparse matrices and from the line accesses for any cached data.

\PIUMA{PIUMA has a high-radix, low-diameter HyperX topology network with all to all connections \cite{Hive2020piuma}. The network is optimized for 8-byte messages as well. The highest level links in the network are optical and support efficient scale-out \cite{Hive2020piuma}. In PIUMA, network bandwidth exceeds local DRAM bandwidth to support higher remote traffics \cite{Hive2020piuma}.} For WMD, we used the DGAS systems to seamlessly communicate using standard load and store operation, and the results shown in the later section would indicate that there is almost no performance hit for moving from a single die (8 blocks) to multi dies on PIUMA.

\section{Optimized Algorithm for WMD}
\subsection{Earth- to Word- Movers Distance}
Algorithm \ref{Algo1} shows how the Sinkhorn distance can be computed for Earth Movers Distance (EMD) \cite{Cuturi:2013:SDL:2999792.2999868}. We adjusted Algorithm \ref{Algo1} \cite{Cuturi:2013:SDL:2999792.2999868} to compute WMD in a way that is simple, easy to follow and implement. 
In the pseudocode, \({d_M^\lambda}(r,c)\) denotes the word movers distance between two documents, \(r\) is the word frequency vector (histogram) of the input/source/query document and \(c\) is the word frequency vector of the target document. If we have multiple target documents, \(c\) would then represent an array of vectors (i.e., a matrix) of word frequencies of all target documents with each column of \(c\), \(c[:j]\) denoting the word frequency vector of target document \(j\). The \(\lambda\) denotes the regularizing entropy parameter and \(M\) denotes the pair-wise transportation cost matrix, storing the euclidean distance among each pair of words in the dictionary. 

\begin{algorithm}[t]
\SetAlgoLined
\KwData{\(M, \lambda, r, c.\)} \KwResult{\({d_M}^\lambda (r, c)\) }
  \(I=(r>0); r=r(I); M=M(I,:); K=exp(-\lambda*M)\) 
  
  Set \(x=ones(length(r),size(c,2))/length(r)\)   
  
  \While{x changes}
  {
  \(x=diag(1./r)*K*(c.*(1./(k^T*(1./x))))\)  
  }
  
  \(u=1./x; v=c.*(1./(k^T*u))\)  
  
  \({d_M}^\lambda (r, c)=sum(u.*((K.*M)*v))\) 
 \caption{Computation of \({d_M}^\lambda (r, c)\) using Sinkhorn-Knopp\cite{Cuturi:2013:SDL:2999792.2999868}.}
 \label{Algo1}
\end{algorithm}

\subsection{Dense vs Sparse WMD Algorithm}
A naive implementation of Algorithm \ref{Algo1} may lead to a dense-compute heavy workload due to the sizes of the dense matrices involved. What creates such a situation is explained below.

We assume that we have one source document and many target documents. We start with a Python implementation of Algorithm \ref{Algo1} with relevant modifications for WMD.

\subsubsection{A Python Implementation}
\begin{figure}[t]
  \centering
  \includegraphics[width=0.5\textwidth]{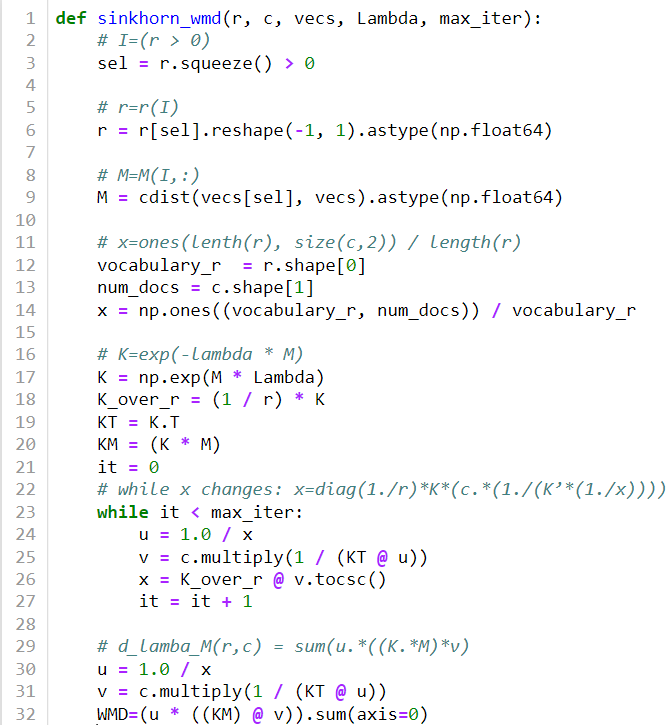}
  \caption{Python Implementation of Algorithm \ref{Algo1}.}
  \label{fig:WMD-python}
  \vspace{-10pt}
\end{figure}

Figure \ref{fig:WMD-python} shows a Python implementation of Algorithm \ref{Algo1}. The sinkhorn\_wmd function takes \(r\), \(c\), \(vecs\), \(Lambda\) and \(max\_iter\) as inputs. Here, \(r\) is a sparse vector of length \(vocabulary\_size\) representing a histogram of word frequencies in the source document. More specifically, \(r[i]\) denotes the normalized count of \(i^{th}\) vocabulary word in the source document. Entries in \(r\) are normalized so that \(sum(r) = 1\). The \(c\) is a sparse \(vocabulary\_size \times num\_docs\) size matrix where \(c[i,j]\) denotes the normalized frequency/count of the \(i^{th}\) word in the \(j^{th}\) target document. The columns of \(c\) are normalized so that sum of normalized word frequency in a particular target document is \(1\), i.e., \(c[:,j] = 1\). 
The \(vecs\) is a \(vocabulary\_size \times word\_embedding\_size\) dense matrix where the \(i^{th}\) row of \(vecs\) provides the word2Vec (or BERT\cite{devlin2018bert} or ElMo\cite{peters2018deep}) word-embedding vector of \(i^{th}\) word in the dictionary/vocabulary. The \(Lambda\) is the regularization parameter and \(max\_iter\) is the maximum number of iterations to find a solution. In an ideal scenario, one would iterate as long as there is any change in the output in the previous iteration, however, in a practice, a cutoff is used to avoid iterating indefinitely. 

In Python, the \(*\) operator works as an element-wise multiplication and \(@\) operator performs the canonical matrix multiplication operation. The Python function sinkhorn\_wmd starts by selecting the nonzero entries of \(r\), i.e., selecting only those words from the vocabulary that appears in the source document. After that, on line 6, it overwrites \(r\) with one that contains only the non-zero entries. Let, \(v\_r\) denotes the number of non-zero entries in \(r\). Next, it computes the euclidean distance (transportation cost) matrix, \(M\) such that \(m(i, j)\) = \(\sqrt{{\sum{{\left(embedding[i] - embedding[j]\right)}^2}}}\), where \(i\) denotes a word that appears in the source document and \(j\) denotes a word that appears in the dictionary. We lazily compute \(M\) based on the nonzero entries of the source document \(r\), instead of computing the full \(vocabulary\_size \times vocabulary\_size\) distance matrix. On line 14, we create the \(x\) matrix of size \(v\_r \times num\_docs\) assuming an equal amount of mass/weight in all points/words. Next, it computes other matrices that need to be computed only once: \(K[i,j]=exp^{Lamda . M[i,j]}\), \(K\_over\_r = (1 / r) * K\), \(KT = K.T\) and \(KM = (K * M)\). Lamda is passed to the function after being negated. Next, it runs the solver loop to iteratively reach close to the optimal solution (i.e., WMD). It computes \(u = 1/x\) (i.e., \(u[i,j]=1/x[i,j]\), then, \(v = c.multiply(1 / (KT @ u))\) and then, \(x = K\_over\_r @  v.tocsc()\). Here, \(c.multiply\) is an element-wise multiplication and \(v.tocsc\) transposes the matrix in sparse format. After the max\_iter number of iterations, the while loop exits. After exiting the while loop, it computes \(u = 1.0 / x\), \(v = c.multiply(1 / (KT @ u))\), followed by \(WMD=(u * ((KM) @ v)).sum(axis=0)\) operation. Here, \(WMD\) is a vector of size \(num\_docs\), where, WMD[i] = Sinkhorn\_distance (src\_doc, target\_doc[i]).

\subsubsection{Dataset}
\label{dataset}
In this paper, we use a precalculated word embeddings as an input that is trained by Google on a very large number of documents scraped from the internet and captures a lot of information about the meanings of words. The dictionary size (i.e., vocabulary\_size) is \(100,000\) words and the word-embedding vector size is \(300\). Therefore, with fp64 numbers, the dictionary size is \(100,000 \times 300 \time 8\)=\(0.24GB\) which does not fit in cache. This precalculated word-embeddings is a subset of the \verb|crawl-300d-2M.vec|\footnote{See for instance \url{https://www.kaggle.com/yekenot/fasttext-crawl-300d-2m\#crawl-300d-2M.vec}} word embedding. We use documents from \verb|dbpedia.train.gz| database\footnote{See for instance \url{https://www.kaggle.com/lotuswhl/dbpediafromfasttext}} as our source and target documents. We use the first \(5000\) documents as our target documents and first \(10\) documents as source/query documents. The properties of the matrices are as follows:
\begin{itemize}
\item \texttt{c}: a sparse matrix with \(100,000\) \(\times\) \(5000\) fp64 elements, holding normalized word frequency in the documents. Approx. \(0.0035\) \% of the total entries are non-zero. c is passed as a CSR. 
\item \texttt {vecs}: a dense matrix with \(100,000\) \(\times\) \(300\) fp64 elements, holding the word embeddings.
\item \texttt{r}: a sparse vector with \(100,000\) elements, holding the word frequency of the input document.
\end{itemize}

\begin{table}[t]
\centering
\vspace{-10pt}
  \caption{Profile of the Python code on an Intel\textregistered{} Xeon\textregistered{} CPU.}
  \scalebox{0.75}
  {    
    \begin{tabular} {|r|l|p{1cm}|}
      \hline
      \multicolumn{1}{|c}{\bf Time} \% & \multicolumn{1}{|c}{\bf Code line} & \multicolumn{1}{|c|}{\bf Potential Kernel} \\ \hline
      0 \% & \texttt{sel = r.squeeze() > 0}  & \\ \hline
      0 \%& \texttt{r = r[sel].reshape(-1, 1).astype(np.float64)}  & \\ \hline
      1.4 \%& \texttt{M = cdist(vecs[sel], vecs).astype(np.float64)}  & Euclidean Distance\\ \hline
      0 \%& \texttt{a\_dim  = r.shape[0]}  & \\ \hline
      0 \%& \texttt{b\_nobs = c.shape[1]}  & \\ \hline
      0 \%& \texttt{x = np.ones((a\_dim, b\_nobs)) / a\_dim}  & \\ \hline
      0 \%& \texttt{K = np.exp(- M * lamb)}  & \\ \hline
      0 \%& \texttt{p=(1 / r) * K}  & \\ \hline
      0 \%& \texttt{it = 0}  & \\ \hline
      0 \%& \texttt{while it < max\_iter}  & Convergence loop\\ \hline
      0 \%& \texttt{u = 1.0 / x}  & \\ \hline
      0 \%& \texttt{KT=K.T}  & Transpose \\ \hline
      91.9 \%& \texttt{v = c.multiply(1 / (KT @ u))}  & \footnotesize{Sparse$\times$Dense$\times$Dense} matrix\\ \hline
      0.1 \%& \texttt{v\_csc=v.tocsc()}  & \\ \hline
      0.5 \%& \texttt{x = K\_over\_r v\_csc}  & Dense$\times$Sparse matrix\\ \hline
      0 \%& \texttt{it += 1}  & \\ \hline
      0 \%& \texttt{u = 1.0 / x}  & \\ \hline
      6.1 \%& \texttt{v = c.multiply(1 / (K.T @ u))}  &  \footnotesize{Sparse$\times$Dense$\times$Dense} matrix\\ \hline
      0 \%& \texttt{return (u * ((K * M) @ v)).sum(axis=0)}  & Dense$\times$Sparse matrix\\ \hline
  \end{tabular} 
  }
  \label{table:sinkhorn_profile}

\end{table}

\subsubsection{Hotspots on Xeon CPU}

We ran the Python code shown in Figure \ref{fig:WMD-python} on a 2-socket ``Cascade Lake'' generation Intel\textregistered{} Xeon\textregistered{} Platinum 8280 system with 28 cores per socket clocked at \(2.70\) GHz. Although the Python code appears to be sequential, it internally used \(48\) threads (observed using \emph{top} command on Linux terminal) and took approximately \(64\) seconds for a source document with \(19\) words (i.e., \(19\) non-zero entries in \(r\)). Those threads were used by Math Kernel Library (MKL)/Blas library calls. Notice that, the Python implementation requires dense matrix multiplication (@) on large arrays \((K.T @ u)\) of size \((100,000 \times 19)\) @ \((19 \times 5000)\) followed by an element-wise multiplication ($*$) by a sparse matrix \(c\) with the resultant dense matrix, \(c.multiply(K.T @ u)\) whereas the sparse matrix \(c\) has only \(173087\) non-zero values out of \((100,000 \times 5000)\) possible entries (i.e., density = \(0.00346174\) \%). The initial Python profiling of the workload using \verb|cProfile| and \verb|line profile| tools is shown in table \ref{table:sinkhorn_profile}. The key computational kernel is the (Sparse * Dense @ Dense) matrix multiplication which takes upto \(98\%\) of the total time. The (Dense @ Sparse) matrix multiplication is insignificant in terms of runtime as shown in Table \ref{table:sinkhorn_profile}. We further profiled the code using Intel's Vtune profiler that highlights that the MKL library calls take the most amount of time in the code.

\begin{algorithm}[hbt]
\SetAlgoLined
\KwData{CSR c[vocabulary\_size][num\_docs],\\ Dense KT[vocabulary]\_size[v\_r],u[num\_docs][v\_r]} 
\KwResult{CSR w[vocabulary\_size][num\_docs]}
  
  \For{i = 0 to c.num\_rows-1s}
  {
  \For{j = c.row\_ptr[i] to c.row\_ptr[i+1] }
    {
        \For{k = 0 to v\_r  }
        {
            w.values[j] += KT[i][k]*u[j][k]
        }
        w.values[j]*=c.values[j] 
    }
  }
    \caption{SDDMM: Sampled Dense Dense Matrix Multiplication}
    \label{SDDMMalgorithm}
\end{algorithm}
\setlength{\textfloatsep}{0pt}

\subsubsection{Dense-heavy to Sparse-heavy}
The dense matrix-multiplication followed by the sparse element-wise multiplication (\(c.multiply(1 /\\ (KT @ u)\)) in the Python implementation is very costly due to the large dense matrices. We converted the entire (sparse * dense @ dense) kernel to a sparse selection of dense matrix multiplication kernel (a SDDMM kernel \cite{hong2019adaptive} shown in Listing. \ref{SDDMMalgorithm}), that performs a dot product only for that row (\(i\)) and a column (\(j\)) for which \(c[i,j]\) is non-zero. That way, instead of doing dense matrix multiplications and then filtering out most of them by a sparse matrix, we first select which dot products are needed using the non-zero \(c[i,j]\)s and then do those dense dot products. The resultant matrix of the SDDMM operation is a sparse matrix which is of the same size as \(c\). This not only eliminates over-computations, but also saves memory bandwidth, intermediate storage and time. Through this algorithmic optimization, the dense-heavy workload is converted to a more efficient sparse workload. Next, we do another matrix multiplication of a dense and sparse matrices \((x = K\_over\_r @ v\_csc)\) using a `non-standard' SpMM (Listing. \ref{SpMMalgorithm}) kernel.

\begin{algorithm}[hbt]
\caption{SpMM: Sparse Matrix Matrix Multiplication}\label{algorithm1}
\SetAlgoLined
\KwData{CSR w[vocabulary\_size][num\_docs],  \\Dense K\_over\_r[vocabulary]\_size[v\_r]} 
\KwResult{Dense x[v\_r][num\_docs]}
  \For{i = 0 to w.num\_rows-1s}
  {
  \For{j = w.row\_ptr[i] to w.row\_ptr[i+1] }
    {
        w\_val = w.values[i][j]
        \For{k = 0 to v\_r  }
        {
            x[i] += K\_over\_r [i][k]*w\_val 
        }
    }
  }
  \label{SpMMalgorithm}
\end{algorithm}
\setlength{\textfloatsep}{0pt}

\subsection{The New SDDMM\_SpMM Kernels} 
\begin{figure*}[t]
  \centering
  \includegraphics[width=0.485\textwidth]{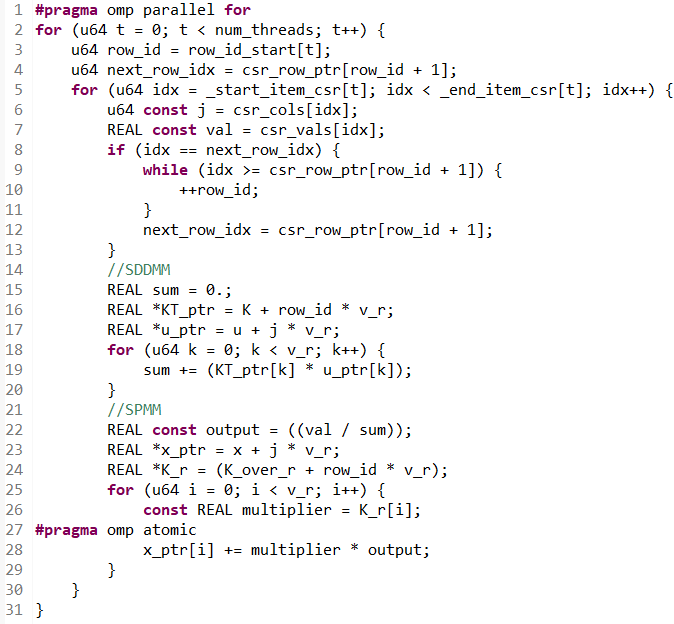}
  \includegraphics[width=0.485\textwidth]{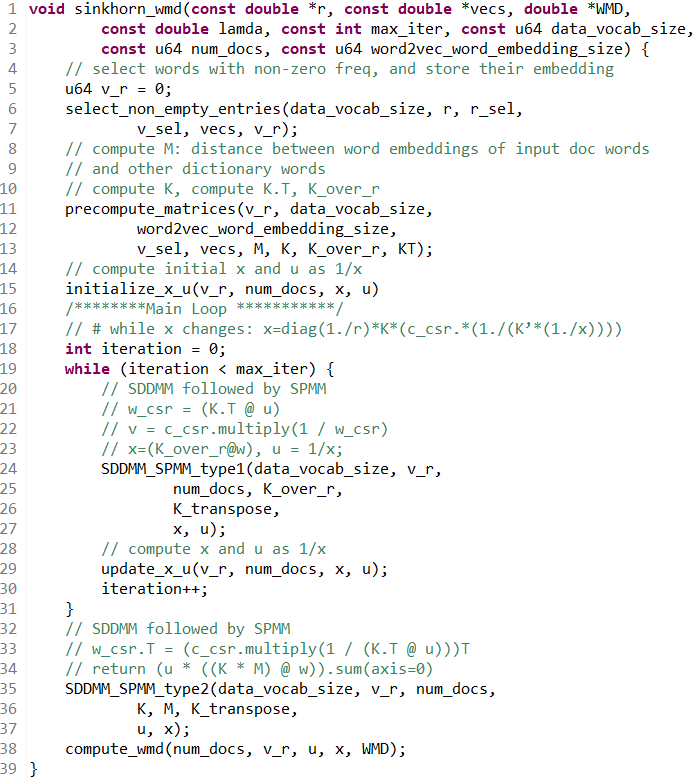}
  \caption{Left: Code snippet of SDDMM\_SpMM kernel, Right: Snippet of the Sinkhorn-Knopp WMD C++ Implementation.}
  \label{fig:SDDMMandSpMM}
  \vspace{-10pt}
\end{figure*}

We fused the SDDMM and SpMM kernels to create a new sparse matrix kernel named SDDMM\_SpMM. \new{To make this possible, we had to fuse the transposition of \(v\_csc\) with SpMM. In other words, we had to do on-the-fly transpose}. In Sinkhorn-WMD, we encounter two slightly different types of SDDMM\_SpMM kernels --- one takes two dense and two sparse matrices (SDDMM\_SpMM\_type1, Line 24 of Figure \ref{fig:SDDMMandSpMM}), and the second one takes three dense and two sparse matrices (SDDMM\_SpMM\_type2, Line 35 of Figure \ref{fig:SDDMMandSpMM}) as input parameters. The benefits of SDDMM\_SpMM are {\bf \(1)\)} it frees us from iterating twice over the CSR rows and column ids (saves memory bandwidth and time), {\bf \(2)\)} the output values from SDDMM can be fed directly to the SpMM and would not need to be stored in the memory (saves memory capacity, write bandwidth for SDDMM and read bandwidth for SpMM). Additionally, data \new{is} transposed on the fly to ensure unit-stride data accesses. \new{This saves storage for \(v\), \(K.T\), \(v\_csc\) and \(v\_csc=v.tocsc()\) conversion step used in the python code, thus, is more memory efficient, reduces cache misses, and saves memory bandwidth}. Figure \ref{fig:SDDMMandSpMM} shows a code snippet of SDDMM\_SpMM. The \(K\_over\_r\), \(M\) matrices are pre-computed once and reused over and over again during the while loop iterations (saving memory bandwidth and compute cycles).

\subsection{Parallelization}
We follow a SPMD (single program multiple data) style parallelization scheme. For better {\bf load-balancing} in the SDDMM\_SpMM kernel, instead of naively dividing the rows across threads (also known as 1D partitioning), we divide the number of non-zeros in the \(c\) matrix evenly among the threads (also known as 2D partitioning). Each thread in parallel determines its starting exploration point inside the CSR using a {\bf binary search} which guarantees an equal work distribution across threads. Since, we parallelize across non-zeros, we need to use atomics (or other conflict resolution techniques) for updating \(x\) in parallel inside SpMM. In addition, we use the following optimizations: 
\begin{itemize}
\item On the fly transpose for unit stride data access.
\item Optimized pointer arithmetic to avoid overheads of address calculation.
\item Loop unrolling, standard optimizations and vectorizations.
\item Function inlining, common sub-expression eliminations.
\end{itemize}
Note that, for the given use case (comparing short texts such as tweets), the \(v\_r\) is quite small (e.g., 19). Luckily, the number of non-zeros is large enough to offer sufficient parallelsim. Hence, we do not parallelize across the \(v\_r\) dimension. In case, where we have large documents or when the number of threads is more than the number of non-zeros, one should use hierarchical threading (omp teams) to exploit parallelism across \(v\_r\) dimension as well and use standard techniques to reduce overheads of atomics. 

We implemented this parallel algorithm using C++ and OpenMP. Figure \ref{fig:SDDMMandSpMM} shows a skeleton of implementation of the SDDMM\_SpMM kernel (left) and Sinkhorn-Knopp WMD algorithm (right). All sub-functions were parallelized using openMP.

\subsection{Theoretical Analysis}
In this section, we show a theoretical runtime analysis of the Sinkhorn\_WMD function shown in Figure. \ref{fig:SDDMMandSpMM}. If the word-embedding vector for each dictionary word is of size \(w\), number of words in the source doc is \(v\_r\), total number of words in the dictionary is \(V\), number of target documents is \(N\), and the number of total non-zeros in the CSR is \(nnz\), then the select\_non\_zero\_entries (Line 6) would take \(O(V + {v\_r \cdot w\over p} + \log p)\) time where \(p\) is the number of threads. The precompute\_matrices (Line 11) would take \(O({V\over p} \cdot v\_r \cdot w + {V\over p} \cdot v\_r + \log p)\) and Initialize\_x\_u (Line 15) would take ~\(O({{N \over p} \cdot v\_r } + \log p)\) time. 
To determine the starting point in CSR, each thread needs to spend \(O(\log V)\) time. Inside the while loop, the SDDMM\_SPMM (both type1 and type2) takes \(O({V\over p} + {nnz\over p} \cdot v\_r +\log p)\), and update\_x\_u takes \(O({N\cdot{v\_r \over p}} + \log p)\). Finally, the compute\_WMD takes \(O(N\cdot {v\_r \over p} + \log p)\). The \(\log p\) terms account for the thread spawning and barrier overheads. Table \ref{tab:cost} shows the asymptotic cost of running our proposed parallel algorithm for computing WMD of one document against N other target documents (1 to N) and the average cost per target document (1 to 1) where the target document has \(v\_x\) unique words and \(t\) is the number of iterations for convergence. 

\begin{table}[t]

  \centering
  \caption{Asymptotic Runtime Cost}
  \vspace{-10pt}
  \resizebox{0.45\textwidth}{!}{
    \begin{tabular}{l|l|l}
    \toprule
          & 1 to N WMD Cost & 1 to 1 WMD Cost \\
          \midrule
    Total cost  & \(O \left({V \cdot v\_r\cdot w  \over p }+ t \cdot {nnz \cdot v\_r \over p}\right)\) & \(O \left({v\_x \cdot v\_r \cdot w  \over p} + t \cdot {v\_x \cdot v\_r \over p}\right)\) \\
    \bottomrule
    \end{tabular}%
    }
  \label{tab:cost}%
\end{table}%

\vspace{-8pt}
\section{WMD on PIUMA}
\label{PIUMA}

In this section we analyze the WMD performance on the PIUMA architecture. \PIUMA{PIUMA provides a primitive programming model with a collection of threads operating on a global shared memory. It supports LLVM C/C++ with and OpenMP extensions to exploit both Single Program Multiple Data (SPMD) and task-based parallelization schemes.}

\subsection{PIUMA programming}
The PIUMA version of the code has been derived from the original C++ implementation. We changed the memory allocation calls to match the PIUMA DGAS memory model. PIUMA provides special features to optimize data allocation according to a user defined pattern, e.g. block cyclic across main memory or scratchpad. We added the PIUMA library header files and linked with the PIUMA runtime library during compilation. We used the STCs for memory allocations and the MTCs to initialize and execute the main kernel in parallel. In addition, we used the following PIUMA specific optimizations:
\setlength{\textfloatsep}{0pt}
\PIUMA{
\begin{itemize}
\setlength{\textfloatsep}{0pt}
\item Caching read-only data (e.g., the word-embedding vecs).
\item Using DMA for data initializations and copying.
\item Using PIUMA builtin\_FP32 transcendental functions to compute sqrt and exp math functions.
\item Scratchpads to store \(x\), \(u\) matrices.
\item PIUMA remote atomics to update \(x\) in parallel throughout the DGAS system.
\item PIUMA builtin barriers and synchronizations to synchronize threads across the DGAS system.
\setlength{\textfloatsep}{0pt}
\end{itemize}
}

\PIUMA
{We used FP64 data type for the input matrices, however, other matrices used FP32 data type. This change was made to enable the use of hardware intrinsics for transcendental. The error for this change in data type was less than \(9.5e^{-07}\) which was within the acceptable error threshold.}

\begin{figure}[tb]
  \centering
  \includegraphics[width=0.32\textwidth]{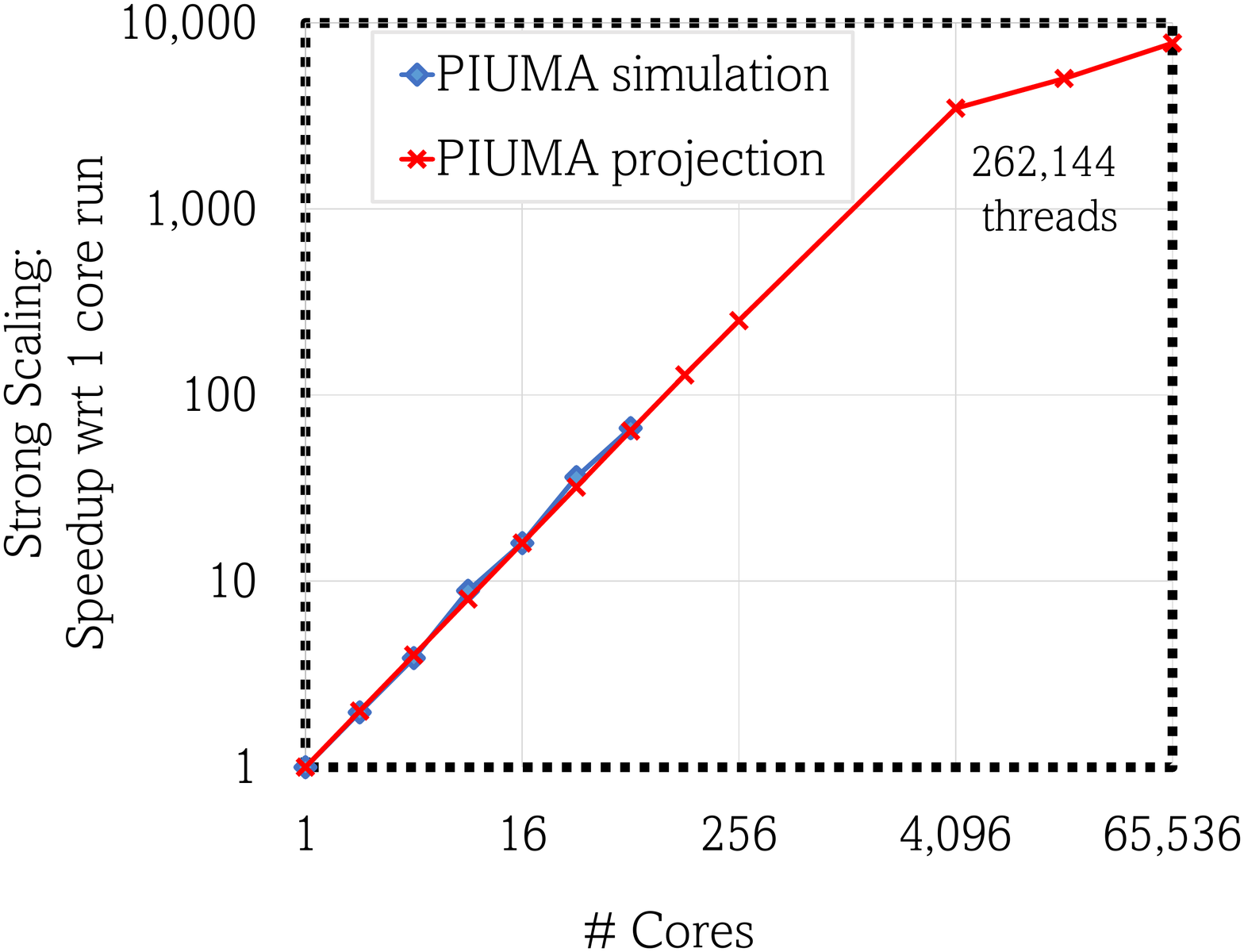}
  \caption[Sinkhorn strong scaling -- simulated and predicted for
    PIUMA.]{Strong scaling -- simulated (up to 64 cores) and
    projected beyond.}
  \label{fig:XeonPUMAPerf}
\end{figure}
\subsection{Simulation Results}
\PIUMA{We have enhanced Sniper \cite{Sniper} to simulate the PIUMA system in a cycle-accurate model as described in \cite{Hive2020piuma}, and we simulated the performance up to \(64\) cores.} \new{Additionally, an FPGA-based RTL system has been used to model up to 1 PIUMA die.} Since simulation time is \(100,000\times\) slower than wall clock, we projected the performance beyond 64 cores using an analytical model. The model uses {{roofline analysis}} to pick the maximum runtime considering the compute, memory, network traffic, and synchronization overhead. \new {We then adjusted the model's projection based on the observed error/deviation from the simulated runtime from \(1\) to \(64\) cores.}

Figure \ref{fig:XeonPUMAPerf} shows the strong scaling performance for the input dataset mentioned in Section \ref{dataset}. The max\_iter was set to 1 to keep simulation time manageable. In Figure \ref{fig:XeonPUMAPerf}, we see a close match between the simulated and predicted time. \PIUMA{The total number of threads in the MTC is 64, and therefore, at 64 cores, we have 4096 threads working at Line 2 of Listing \ref{fig:SDDMMandSpMM}. With 5000 target documents and 100,000 vocabulary words, beyond \(2048\) cores (131k threads), the work (nnz=173,087) becomes less compared to the number of threads. At this point, the barrier costs become significant.} \new{Beyond 1 die (8 cores), PIUMA cores start utilizing network bandwidth to access data distributed across all memory controllers. Therefore, simulation on 64 cores (8 dies) shows native (seamless) scale-out efficiency of PIUMA.}

\begin{figure}[tb]
 \centering
  \includegraphics[width=0.4\textwidth]{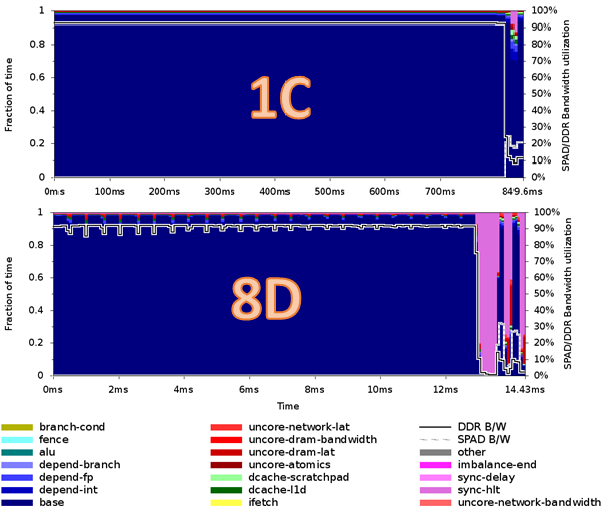}
  \caption {CPI plots for 1 iteration on 1 core and 8 dies (simulated on PIUMA). The plot provides at-a-glance view of the system performance, capturing aggregate metrics of memory utilization (DDR Bandwidth), and computational activities of the main cores and uncore units.}
          \label{fig:sinkhorn-cpi}%
\end{figure}%

\PIUMA{Figure \ref{fig:sinkhorn-cpi} shows CPI (Cycles Per Instruction) plots \cite{Sniper} generated by the Sniper simulation of the sinkhorn-WMD workload for single iteration on 1C (core) and 8D (8 Dies, 4096 MTC threads). \new{"CPI stacks are a first-order method for understanding the causes of performance loss in an out-of-order processor"} \cite{Sniper}. The CPI plot breaks up an application’s execution time into a number of components normalized to cycles per instruction. The first dark blue portion shows performance during the euclidean distance computation which is mainly a dense compute portion. The second portion of the CPI plot shows the convergence loop which is mostly the iterative SDDMM\_SpMM kernels followed by copies and inverse operations. This part is a mix of sparse and dense compute. The dark blue portion of the plots indicates that all pipelines are full executing instructions. The black line on top shows the effective memory bandwidth. The dark blue touching the 100\% indicates that the code is running at its peak of the number of instructions that can be executed per cycle. The black line close of 90\% percentage indicates that the implementation is achieving over 90\% of the system’s bandwidth depending on the given the mix of read and write traffic. The pink portion shows synchronization delays. 
}

\begin{figure}[t]
  \centering
 
  \includegraphics[width=0.4\textwidth]{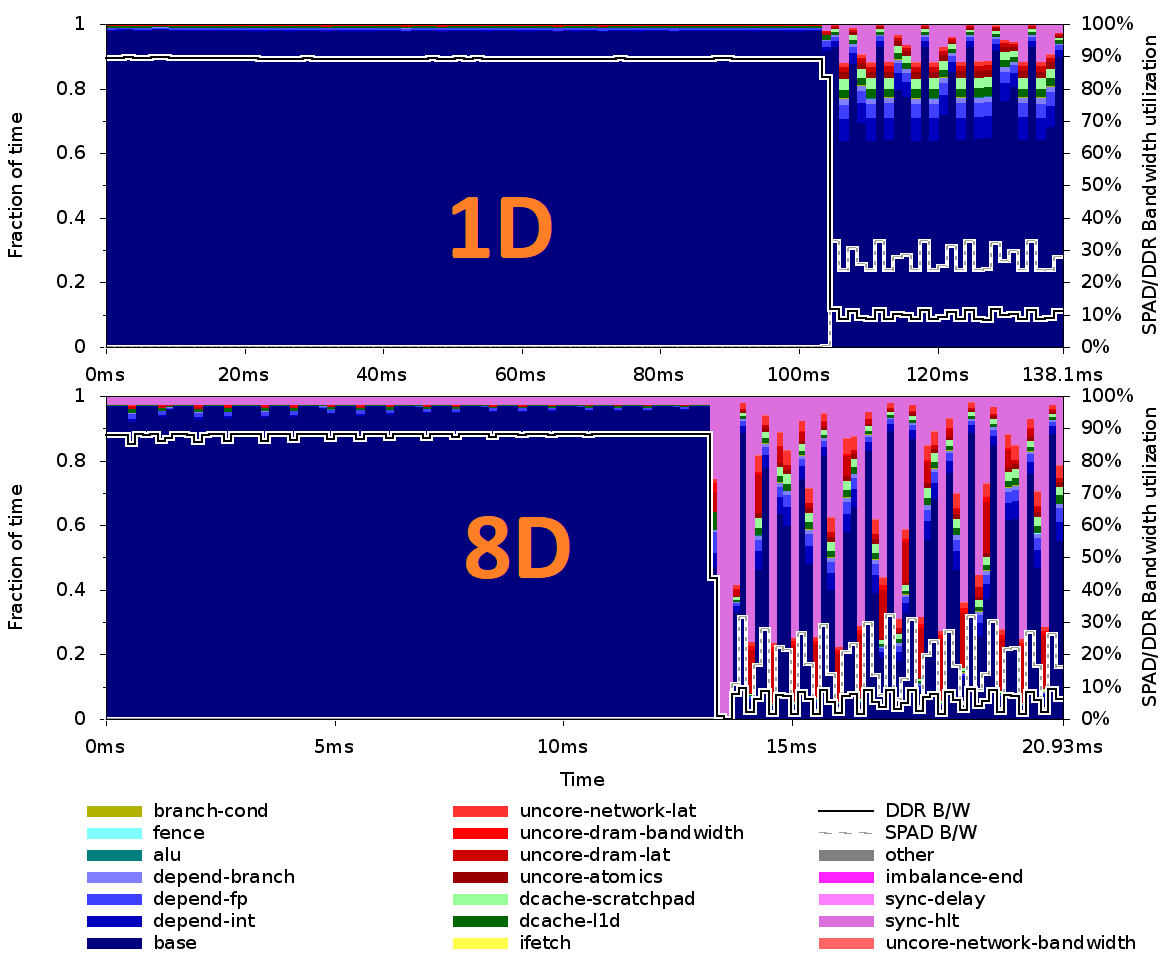}
  \caption{CPI plots for 15 iterations simulated on 1D and 8D of PIUMA.}
  \label{fig:sinkhorn-cpi-15t}%
\end{figure}

Figure~\ref{fig:sinkhorn-cpi-15t} shows CPI plots for max\_iter=\(15\). \PIUMA{On PIUMA, most of the time is spent in the euclidean distance computation part. On the other hand, the sparse portion is relatively faster. This is opposite to what we saw in Table \ref{table:sinkhorn_profile} for Xeon. Note that, PIUMA cores are scalar and currently support low arithmetic intensity (less than 2 flops-per-byte). On the other hand, Xeon CPUs have arithmetic intensity in the range of 20s for FP64 type. \new{An Intel Advisor analysis on the CPU code shows that the loops inside the euclidean distance computation were vectorized by the compiler on Xeon and the speedup through vectorization was over 5x. PIUMA cores, being scalar does not have this benefit.} Thus, it is expected that the dense euclidean distance computation part would take longer on PIUMA.
}

\PIUMA{
To {{summarize}}, we leverage the following features of PIUMA to port and strong scale the Sinkhorn-WMD algorithm: 1) 8-byte access granularity 2) hardware intrinsics for FP32 transcendental 3) DMA for efficient data initialization and copy from/to remote memory 3) Distributed global address space to store the data in a distributed manner, yet access using normal load/store without worrying much about access latency \new{(native scale-out)} 4) the builtin SPMD style programming, 5) selective caching to cache limited data 6) use of SPAD for the data with atomics and 5) remote atomics to resolve conflicts. 
}

\PIUMA{
The { lessons learned} are 1) it is fairly easy to port algorithm to PIUMA, 2) SPMD style parallel implementation scales well up to a large number of threads before hitting plateau, 3) although PIUMA is not designed for dense workloads, it can run close to its peak performance for a workload that is a mix of dense and sparse kernels (Sinkhorn-WMD).} 

In the next section, we show this algorithm's performance on Xeon CPUs.

\section{WMD on Xeon CPU}
\vspace{-3pt}
\begin{table}[b]
  \centering
  \caption{System Specifications}
  \vspace{-5pt}
 \resizebox{0.4\textwidth}{!}
  {
    \begin{tabular}{cc}
    \toprule
    \textbf{Platforms} & \textbf{CLX0} \\
    \midrule
    Model & Intel(R) Xeon(R) Platinum 8280 CPU @ 2.70 GHz \\
    Cpu MHz & 1800 \\
    L1d &  32 KiB \\
    L2  &  1024 KiB \\
    L3  & 39.4 MiB \\
    MemAvailable &  190 GiB \\
    \#Cores per socket & 28 \\
    \#Numa sockets & 2 \\
    \bottomrule
    \end{tabular}%
    \vspace{-15pt}
    }
  \label{tab:sys-spec}%
\end{table}%

We implemented the parallel Sinkhorn-WMD algorithm in C++/OpenMP and compiled the program using the Intel icc version 19.0.2.187 (gcc version 4.8.5 compatibility) with the following compiler flags: \texttt{-O3 -fopenmp  -xHost -g -restrict -std=c++11 -finline -unroll -ansi-alias -qopt-subscript-in-range}. We used a state-of-the-art Intel Xeon system code-named Cascade Lake (CLX). Table \ref{tab:sys-spec} shows the machine specifications. The operating system was CentOS Linux 7. \new{The optimization report shows that the compiler was able to vectorize the loops called inside the euclidean distance computation with AVX\_512 ISA and used shuffles, inserts, extracts, FMA. Other optimizations were threading, unroll and jam, and unroll.}

\begin{figure}[t]
  \centering
  \resizebox{0.52\textwidth}{!}
  {
  \includegraphics[width=0.25\textwidth]{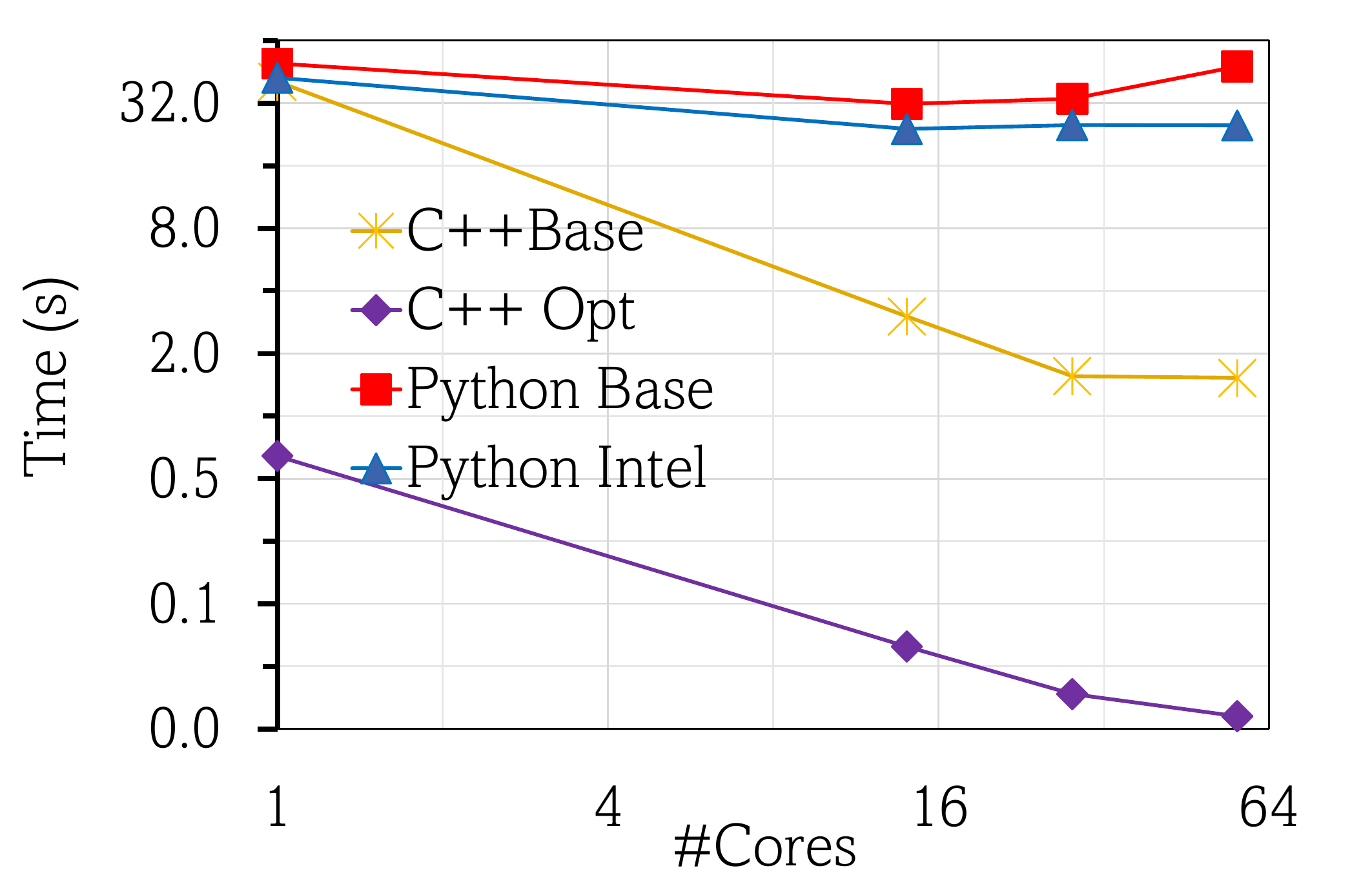}
  \includegraphics[width=0.25\textwidth]{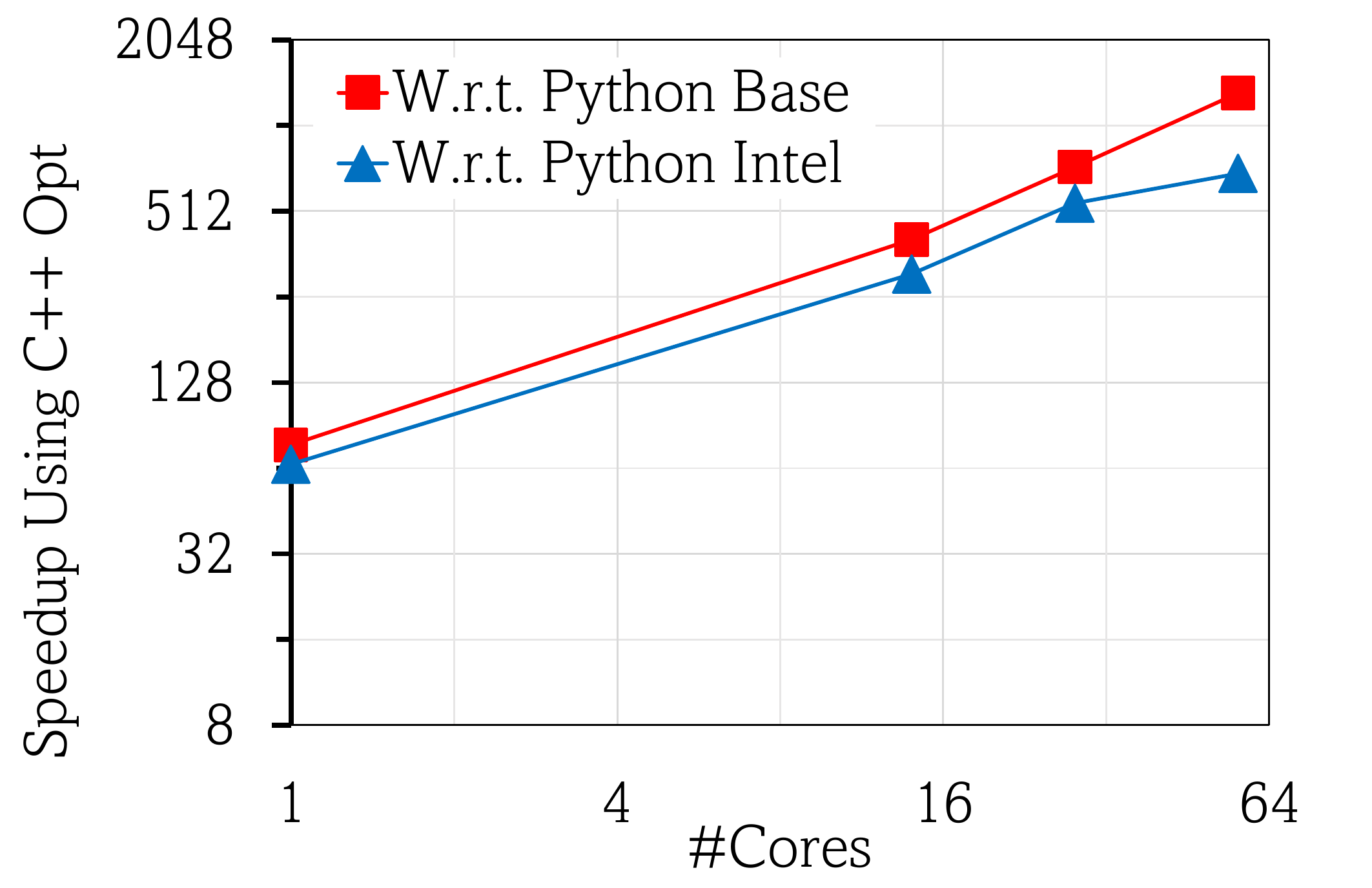}
  }
 \vspace{-8pt}
\caption{1) Runtime of all CPU implementations 2) Speedup wrt. python code}
\label{fig:multisockettime}
\end{figure}

\paragraph{Python vs C++/OpenMP Implementation} First, we used the same \(19\)-word document as the source input from Section \ref{dataset}. For this input, although, the Python code takes around \(47\) sec on the CLX system, the C++/OpenMP implementation takes only \(0.04\) second {\textbf(\(1331\times\) faster)}. Initially, we used a default python installation (Python 3.6.2). Then we used an Intel optimized version (Python 3.7.10, \url{https://github.com/intel/optimized-models/tree/master/pytorch/dlrm}) and with that, the strong scaling improved and the speedup number wrt. C++ code reduced to \(694\times\). The key factors behind this two orders of magnitude speedup are 1) Python being an interpreted language and hence slower, 2) The Python library was unable to exploit the SDDMM kernel. Instead, it naively computed a sparse x dense x dense multiplication and incurred the associated overhead of extra compute, bandwidth, and storage, 3) Frequent creation and transposition of the sparse matrix \new{and associated intermediate storage (for example, the C++ version does need to create \(K.T\), \(v\) and \(v\_csc\) that the Python version uses)}, and finally, 4) Not being able to fuse SDDMM and SpMM and hence has the additional overhead of extra compute and storage. The C++/OpenMP implementation eliminates all those inefficiencies. Figure \ref{fig:multisockettime} shows the trend. We also implemented a baseline C++ translation of the Python code (without the SDDMM kernel) and that is over \(42\times\) slower than the optimized C++ implementation with the SDDMM-SpMM kernel, however, \(32\times\) faster than the Python counter-part.

\begin{figure}[tp]
  \centering
  \resizebox{0.32\textwidth}{!}
  {
  \includegraphics[width=0.4\textwidth]{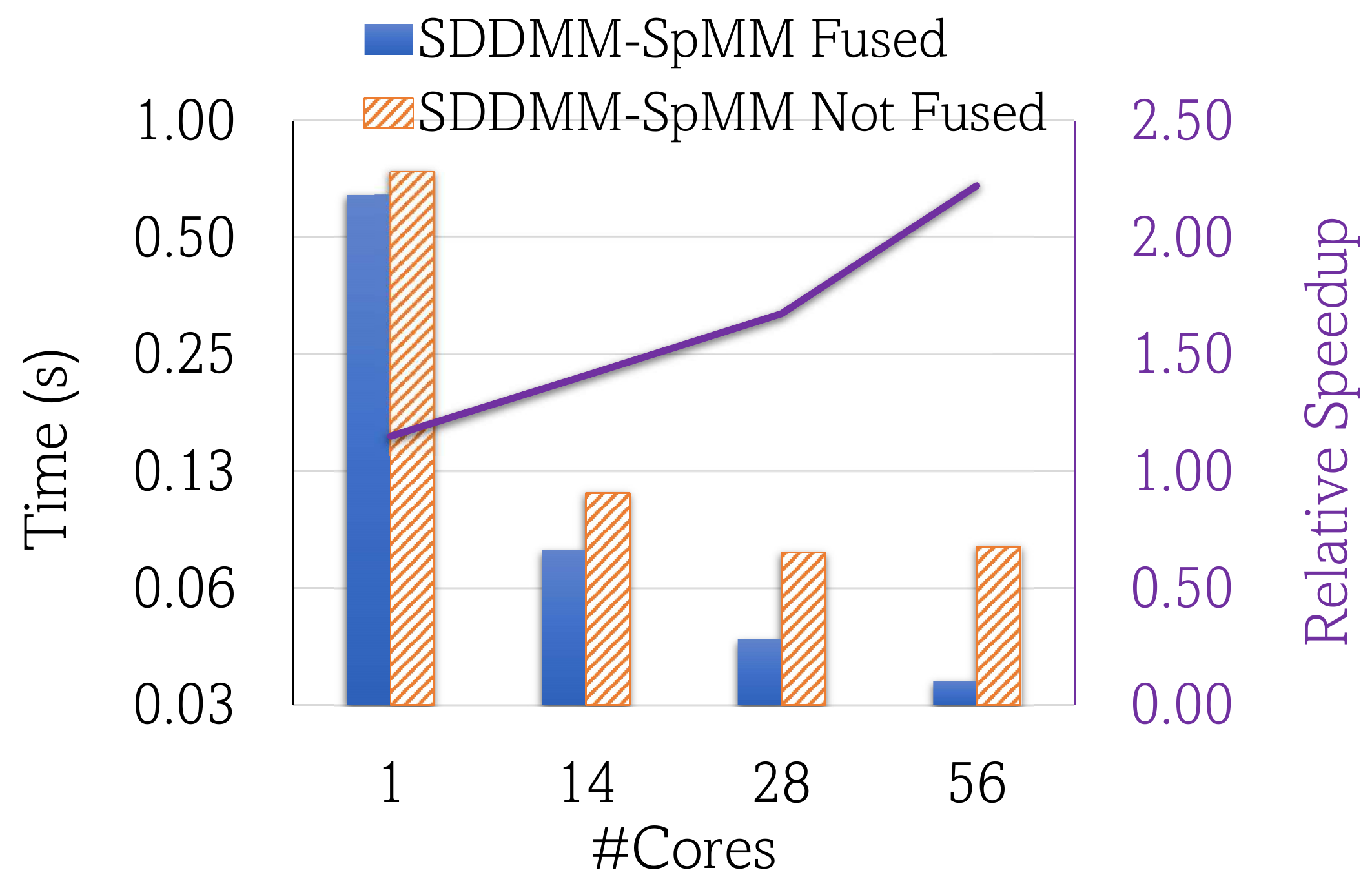} 
  }
\caption{Speedup from Fusion (v\_r=19).}
\label{fig:FusionCLX}
\end{figure}


\paragraph{Speedup From Fusion}
  Figure \ref{fig:FusionCLX} shows the benefits of SDMM-SpMM kernel fusion for the Sinkhorn-WMD algorithm on the C++ implementation. Overall, the speedup ranges from \(1.15\times\) to \(2.22\times\), and the speedup increases with the number of cores. This speedup comes from converting memory write/read operations to register read/write operations and from the elimination of extra loop and storage.

\begin{figure}[t]
 \vspace{-5pt}
  \centering
  \resizebox{0.5\textwidth}{!}
  {
  \includegraphics[width=0.25\textwidth]{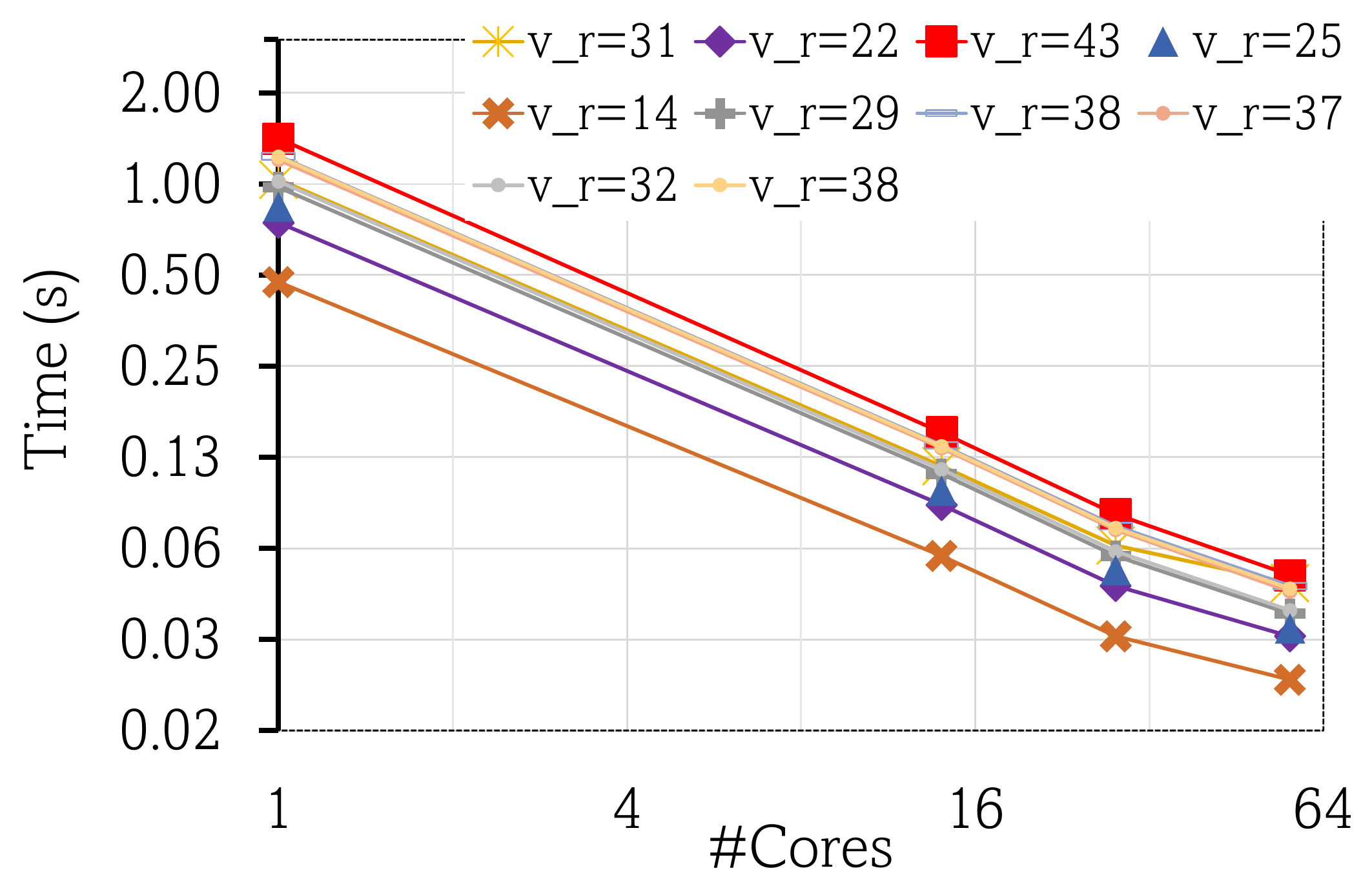}
   \includegraphics[width=0.25\textwidth]{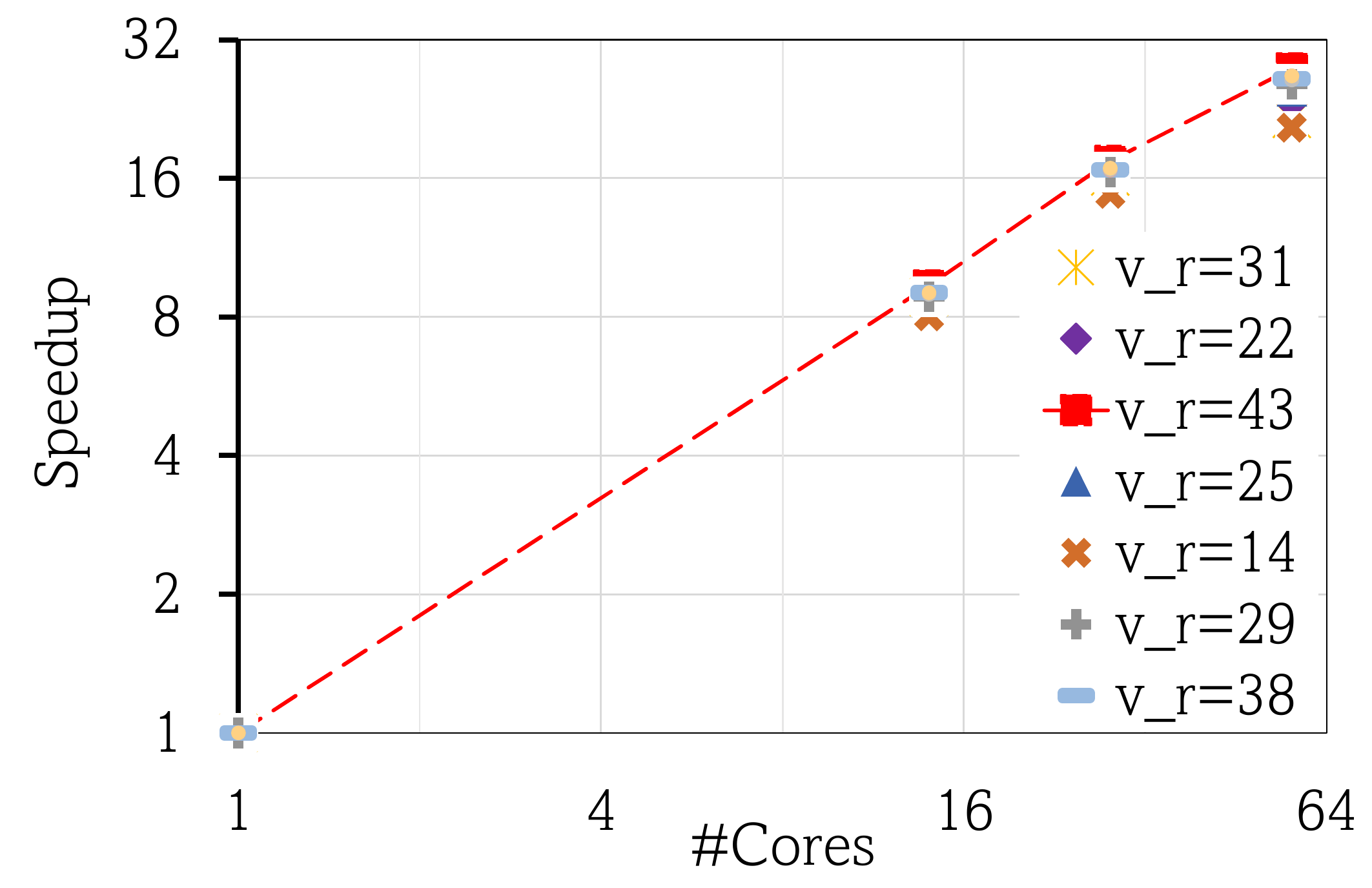}
  }
  \vspace{-10pt}
\caption{Strong scaling on multiple source documents.}
\label{fig:multiTimeCLX}
\end{figure}

\paragraph{Strong Scaling and Speedup} 
Figure \ref{fig:multiTimeCLX} shows the strong scaling for the parallel Sinkhorn-WMD algorithm on multiple source documents if run at once. In the Figure, \(v\_r\) stands for the number of words in the source document. The maximum speedup obtained was for \(v\_r=43\), the highest word document in the batch and it was about \(28\times\) speedup on two-sockets (\(56\)-cores) of CLX. The \(v\_r=14\) had the worst speedup because it is the shortest document. For a bandwidth-bound kernel, if a single CLX core can draw 7.5 GB/s, the expected maximum speedup is \(27.3\times\) (full system bandwidth is 210 GB/s). Thus, we reached close to the peak.
\vspace{-5pt}
\section{Conclusion}
In this paper, we present a new parallel algorithm with better asymptotic bounds to compute the Word Movers Distance that transforms the dense-heavy EMD approximation algorithm into a sparse-heavy algorithm using a class of new SDDMM\_SpMM kernels. Use of EMD for WMD, sparsification of EMD, Parallelization with equal nnz partitioning, new fused kernels, and porting on new architecture are all novelties of this work. We describe how we ported this algorithm to the new PIUMA DGAS architecture using its unique features such as DMAs, SPAD, remote atomics, FP32 hardware intrinsics, and selective caching. We also optimize it on Xeon. Our analysis shows that the presented algorithm reaches close to peak performance on both and shows good strong scaling.

We are making the code public. Thus, our work would 1) help ML developers by providing a faster (\(700\times\)) algorithm to compute WMD in their data science applications, and 2) provide valuable insights into PIUMA's performance for workloads that include a mix of dense and sparse kernels. Furthermore, although this paper focuses on Word-movers distance, the algorithmic improvement ideas presented in this paper are also applicable to any other applications that use Earth-movers distance as a distance metric. Thus, our work has broader applicability as well.


\vspace{-5pt}
\bibliographystyle{IEEEtran}
\bibliography{IEEEabrv,sample-base}

\begin{thebibliography}{10}
\providecommand{\url}[1]{#1}
\csname url@samestyle\endcsname
\providecommand{\newblock}{\relax}
\providecommand{\bibinfo}[2]{#2}
\providecommand{\BIBentrySTDinterwordspacing}{\spaceskip=0pt\relax}
\providecommand{\BIBentryALTinterwordstretchfactor}{4}
\providecommand{\BIBentryALTinterwordspacing}{\spaceskip=\fontdimen2\font plus
\BIBentryALTinterwordstretchfactor\fontdimen3\font minus
  \fontdimen4\font\relax}
\providecommand{\BIBforeignlanguage}[2]{{%
\expandafter\ifx\csname l@#1\endcsname\relax
\typeout{** WARNING: IEEEtran.bst: No hyphenation pattern has been}%
\typeout{** loaded for the language `#1'. Using the pattern for}%
\typeout{** the default language instead.}%
\else
\language=\csname l@#1\endcsname
\fi
#2}}
\providecommand{\BIBdecl}{\relax}
\BIBdecl

\bibitem{WMD}
\BIBentryALTinterwordspacing
M.~J. Kusner, Y.~Sun, N.~I. Kolkin, and K.~Q. Weinberger, ``From word
  embeddings to document distances,'' in \emph{Proceedings of the 32Nd
  International Conference on International Conference on Machine Learning -
  Volume 37}, ser. ICML'15.\hskip 1em plus 0.5em minus 0.4em\relax JMLR.org,
  2015, pp. 957--966. [Online]. Available:
  \url{http://dl.acm.org/citation.cfm?id=3045118.3045221}
\BIBentrySTDinterwordspacing

\bibitem{word2vec}
T.~Mikolov, K.~Chen, G.~Corrado, and J.~Dean, ``Efficient estimation of word
  representations in vector space,'' \emph{arXiv:1301.3781}, 2013.

\bibitem{10.1007/978-3-030-14799-0_11}
N.~Franciscus, X.~Ren, J.~Wang, e.~N.~T. Stantic, Bela, F.~L. Gaol, T.-P. Hong,
  and B.~Trawi{\'{n}}ski, ``Word mover's distance for agglomerative short text
  clustering,'' in \emph{Intelligent Information and Database Systems}.\hskip
  1em plus 0.5em minus 0.4em\relax Cham: Springer International Publishing,
  2019, pp. 128--139.

\bibitem{brokos2016using}
G.-I. Brokos, P.~Malakasiotis, and I.~Androutsopoulos, ``Using centroids of
  word embeddings and word mover's distance for biomedical document retrieval
  in question answering,'' \emph{arXiv preprint arXiv:1608.03905}, 2016.

\bibitem{zhang2016building}
M.~Zhang, Y.~Liu, H.~Luan, M.~Sun, T.~Izuha, and J.~Hao, ``Building earth
  mover's distance on bilingual word embeddings for machine translation,'' in
  \emph{Thirtieth AAAI Conference on Artificial Intelligence}, 2016.

\bibitem{wu2018word}
L.~Wu, I.~E. Yen, K.~Xu, F.~Xu, A.~Balakrishnan, P.-Y. Chen, P.~Ravikumar, and
  M.~J. Witbrock, ``Word mover's embedding: From word2vec to document
  embedding,'' \emph{arXiv preprint arXiv:1811.01713}, 2018.

\bibitem{tashu2018pair}
T.~M. Tashu and T.~Horv{\'a}th, ``Pair-wise: Automatic essay evaluation using
  word mover's distance.'' in \emph{CSEDU (1)}, 2018, pp. 59--66.

\bibitem{pele2009fast}
O.~Pele and M.~Werman, ``Fast and robust earth mover's distances,'' in
  \emph{12th International Conference on Computer Vision}.\hskip 1em plus 0.5em
  minus 0.4em\relax IEEE, 2009.

\bibitem{Cuturi:2013:SDL:2999792.2999868}
\BIBentryALTinterwordspacing
M.~Cuturi, ``Sinkhorn distances: Lightspeed computation of optimal transport,''
  in \emph{Proceedings of the 26th International Conference on Neural
  Information Processing Systems - Volume 2}, ser. NIPS'13.\hskip 1em plus
  0.5em minus 0.4em\relax USA: Curran Associates Inc., 2013, pp. 2292--2300.
  [Online]. Available: \url{http://dl.acm.org/citation.cfm?id=2999792.2999868}
\BIBentrySTDinterwordspacing

\bibitem{Knight:2008:SAC:1404637.1404647}
\BIBentryALTinterwordspacing
P.~A. Knight, ``The sinkhorn-knopp algorithm: Convergence and applications,''
  \emph{SIAM J. Matrix Anal. Appl.}, vol.~30, no.~1, pp. 261--275, Mar. 2008.
  [Online]. Available: \url{http://dx.doi.org/10.1137/060659624}
\BIBentrySTDinterwordspacing

\bibitem{Hive2020piuma}
S.~Aananthakrishnan, N.~K. Ahmed, V.~Cave, M.~Cintra, Y.~Demir, K.~D. Bois,
  S.~Eyerman, J.~B. Fryman, I.~Ganev, W.~Heirman \emph{et~al.}, ``Piuma:
  Programmable integrated unified memory architecture,'' \emph{arXiv preprint
  arXiv:2010.06277}, 2020.

\bibitem{orlin1993faster}
J.~B. Orlin, ``A faster strongly polynomial minimum cost flow algorithm,''
  \emph{Operations research}, vol.~41, no.~2, pp. 338--350, 1993.

\bibitem{Li:2019:CES:3308558.3313397}
\BIBentryALTinterwordspacing
C.~Li, J.~Ouyang, and X.~Li, ``Classifying extremely short texts by exploiting
  semantic centroids in word mover's distance space,'' in \emph{The World Wide
  Web Conference}, ser. WWW '19.\hskip 1em plus 0.5em minus 0.4em\relax New
  York, NY, USA: ACM, 2019, pp. 939--949. [Online]. Available:
  \url{http://doi.acm.org/10.1145/3308558.3313397}
\BIBentrySTDinterwordspacing

\bibitem{hong2019adaptive}
C.~Hong, A.~Sukumaran-Rajam, I.~Nisa, K.~Singh, and P.~Sadayappan, ``Adaptive
  sparse tiling for sparse matrix multiplication,'' in \emph{Proceedings of the
  24th Symposium on Principles and Practice of Parallel Programming}.\hskip 1em
  plus 0.5em minus 0.4em\relax ACM, 2019, pp. 300--314.

\bibitem{gale2020sparse}
T.~Gale, M.~Zaharia, C.~Young, and E.~Elsen, ``Sparse gpu kernels for deep
  learning,'' \emph{arXiv preprint arXiv:2006.10901}, 2020.

\bibitem{rahman2020fusedmm}
M.~Rahman, M.~H. Sujon, A.~Azad \emph{et~al.}, ``Fusedmm: A unified sddmm-spmm
  kernel for graph embedding and graph neural networks,'' \emph{arXiv preprint
  arXiv:2011.06391}, 2020.

\bibitem{devlin2018bert}
J.~Devlin, M.-W. Chang, K.~Lee, and K.~Toutanova, ``Bert: Pre-training of deep
  bidirectional transformers for language understanding,'' \emph{arXiv preprint
  arXiv:1810.04805}, 2018.

\bibitem{peters2018deep}
M.~E. Peters, M.~Neumann, M.~Iyyer, M.~Gardner, C.~Clark, K.~Lee, and
  L.~Zettlemoyer, ``Deep contextualized word representations,'' \emph{arXiv
  preprint arXiv:1802.05365}, 2018.

\bibitem{Sniper}
\BIBentryALTinterwordspacing
T.~E. Carlson, W.~Heirman, S.~Eyerman, I.~Hur, and L.~Eeckhout, ``An evaluation
  of high-level mechanistic core models,'' \emph{ACM Trans. Archit. Code
  Optim.}, vol.~11, no.~3, Aug. 2014. [Online]. Available:
  \url{https://doi.org/10.1145/2629677}
\BIBentrySTDinterwordspacing

\end{thebibliography}

\end{document}